\documentstyle[12pt,epsfig]{article}
\begin{document}
\baselineskip 22pt plus 2pt
\begin{center}
{\bf MECHANICAL ASPECT OF CHIRALITY} \\

{\bf AND ITS BIOLOGICAL SIGNIFICANCE}
\vspace{0.5cm}

{\bf G. Gilat} \\

Department of Physics \\
Technion, Haifa 32000, Israel
\vspace{0.5cm}

{\bf Abstract}
\end{center}
\vspace{0.5cm}

Chirality is not just a structural artifact in biology but it may  
provide for a genuine biological advantage. This is due to the 
phenomenon of chiral interaction (CI) which is described here for 
mechanical-chiral devices. 
The main mechanical feature of chiral interaction is its mode
of selecting 
one direction of rotation out of two possible and opposite ones.
For example, a given chiral device such as a rotating water sprinkler,
rotates in one direction. What does rotate in the opposite direction 
is the mirror of this given sprinkler. This mode of operation indicates
space-time (PT) invariance which causes it to be also time-irreversible.
This also causes a chiral device
to become non-ergodic 
on microscopic level. This prevents certain chiral systems from 
readily reaching thermal equilibrium, and causes the system to act 
non-ergodically, which is crucial for living systems as well as for 
molecular evolution.
\pagebreak

\noindent {\bf I. Introduction} \\

The phenomenon of structural chirality of crystals and molecules 
has been
recognized since the early 19th century when Arago$^1$ and Biot$^2$
did demonstrate the effect of optical activity in quartz crystals. 
Louis Pasteur$^3$ was the first to observe chirality on a 
molecular level and specified it as ``dissymmetry''. The term 
``Chirality'' was first proposed by Kelvin$^4$, who also defined this
concept as a property of any object that cannot superimpose, or 
overlap, completely its mirror image.
It is well known nowadays that most biological molecules consist only of
one out of two possible enantiomers, e.g., left-handed (L) amino-acids
or right-handed (D) sugars. This phenomenon leads to an interesting
question concerning the origin of such a selection 
and there exist several speculations that try to solve this
enigma. A considerably more constructive question to be asked is: ``why
are the molecules of life chiral?'', or ``is there any biological 
advantage in their chiral nature when compared to achiral molecules?''
And the answer to be given here is: ``Yes'', and this is regardless of
their being L or D. The source of such an advantage comes from a specific
type of {\bf interaction} that exists between various mechanical devices
and different media such as flow of air or water and even light 
radiation. What is special about this interaction is the presence
of chiral structure in these devices which makes their mode of 
operation quite different from other interactions which are based
on achiral objects such as the Newtonian mass point. Such an 
interaction is to be labeled ``chiral interaction'' (CI) and it has
already been described and treated in several publications.$^{5-7}$ \
It is interesting to note that this phenomenon of chirality is 
largely being overlooked in classical mechanics, and only a few
physicists are aware of it.
\vspace{0.5cm}

\noindent {\bf II \ Chiral Interaction in Mechanical Devices} \\

As mentioned above
Chiral interaction is not limited to molecular structure only but there
exist various mechanical chiral devices that function according to the
same principle. The most
spectacular example is the rotating windmill. When wind blows at the rotors
of a mill it ``knows'' immediately in which direction to rotate, clock-
or anti-clockwise. If the windmill, in particular its vanes, were
symmetric with respect to their axis of rotation, the 
mill would not be able ``to make up its mind'' in which direction to
rotate. The shape of the vanes that come in contact with the wind is
designed to break the L-D symmetry in order to choose {\bf one} specific
sense of rotation out of {\bf two} possible ones. In other words, the
shape of the vanes where they come in contact with the wind, 
is {\bf chiral}. Another simple mechanical chiral device is the
rotating water sprinkler, or the wind propeller. 
The next example, shown in Fig. 1, is somewhat
more sophisticated, and it depends on a different mode of chirality. 
This device is a simple variant of the Crookes' radiometer. The active
medium in this case is light radiation and the element of chirality 
consists of two different colors on both sides of the 
rotating blades, being black and silver, respectively. This is a special
example of a physical rather than a geometric chirality.
Physical chirality$^{7,8}$ is presented by a chiral distribution 
of a physical property rather than of a chiral geometric shape.
Physical chirality differs from a geometric one in its capability
of interacting with various media surrounding it.
In the case of this special example of a variant of the ordinary Crookes'
radiometer, the physical distribution of the black and silver colors
on the blades represents a large difference in the light absorption 
coefficient
of the blades. The silver side reflects back the light, whereas 
the black side absorbs the light and therefore becomes 
warmer in comparison to the silver
one. This causes the air at the black side to become heated
and as a result it expands and pushes back the black blade
which ends up in rotating the device
in {\bf one} preferred direction out of {\bf two} possible ones, that 
is, in the direction of the black side of the blade. The selection of 
the sense of rotation of the blades is made
by the variance of colors on the blades and their interaction with 
light. The physical chirality$^{7-8}$ here is represented 
by the distribution
of the optical absorption coefficient on the blades and {\bf not}
by their geometric shape. \\

So far, all the examples presented here are of mechanical nature, i.e.,
the effect of chiral interaction (CI) results in a mechanical rotation
in one preferred direction out of two possible ones around a 
given axis of
the device. 
This is so because the source of the interaction, i.e. the medium,
usually is external to the chiral device. In the case of an electric
device which generates a static current flowing in one preferred 
direction out of two possible ones, the source of the interaction may
be embedded within the device. This is the case, for instance, of an 
electric cell which consists of two different electrodes coming in 
contact with an electrolyte. It is obvious that in order to reverse 
the direction of the current it is necessary to interchange the two 
electrodes with one another, but this does not necessarily require any 
chiral operation. This is so because the source of the current flow
is internal, so that the structure of the device can be 
designed to be completely
symmetric, as is the case of a cylindric battery. In the case of an
electric thermocouple, the operation can still be regarded as CI 
since the source of the interaction, i.e. the temperature difference,
is external to the device.  \\
 
To summarize the main features of CI in mechanical-devices let us notice
that in all these examples there exists a specific medium with which
the chiral device is interacting and this always happens at an interface
separating the device from the active medium. The physical chirality is
{\bf built} into this very interface. CI is a process by which energy
is transferred from the active medium into the chiral device 
which causes a rotational motion, being usually of mechanical 
nature. The most significant aspect of the chiral 
interaction process is
its mode of selecting only {\bf one} direction of rotation out of 
{\bf two} possible ones, which is to be attributed to the {\bf chiral}
nature of the device. The {\bf mirror image} of the given chiral device,
interacting with the same medium, does produce the same rotational motion
in the {\bf opposite} direction. This is to be regarded as a main 
feature of chiral interaction. \\

The effect of CI on a molecular level is less recognized in 
comparison to that of macro-chiral devices. The main reason for this is
that CI occurs mainly within the chiral system, or molecule, in the form
of a small perturbation which is not easy to detect experimentally.
Much more recognizable are the physical effects associated with molecular
chiral structure, such as optical activity and related effects. These are 
to be regarded as ``chiral scattering'', rather than CI,
since the observable effect concerns
the polarized light being scattered away from the chiral molecule 
rather than its effect on
the molecule itself which is the chiral 
interaction$^{7}$. \\

A physical model of chiral interaction (CI) in soluble proteins and 
amino-acids has already been developed and described in 
detail in several\\
publications$^{5-7,9-10}$. The description here contains only a few main 
features of this model.  The active medium in this model consists of 
random motion of ions throughout the solvent, being mostly regular
water. The chiral element that interacts with these ions is an electric
dipole moment that exists in the protein structure. This interaction 
causes the moving ion to be deflected away from its original track of
motion, which creates a continuous perturbation along the $\alpha$-helix
of which the proteins consist, and this perturbation moves along the 
helix in {\bf one} preferred direction out of {\bf two} possible ones.
This is an abbreviated description of the model of the CI that occurs
in soluble proteins. A 
more detailed description appears in earlier publications.$^{5-7,9-11}$ 
The perturbation 
resulting from this CI is of electric nature, rather than mechanical
one. Another interesting aspect of this CI is that it happens at
an interface separating the interior of the protein molecule from the 
solvent and this is due to the globular structure of the soluble 
protein. It is well known that all soluble proteins become 
globular before they can function as enzymes$^{12}$. \\

As mentioned above chiral interaction is not easy to observe 
experimentally on a molecular
level due to the smallness of this effect. Nevertheless, there exists
a certain strong supporting evidence owing to an experiment performed 
by Careri et al.$^{13}$. This experiment concerns the effect of 
dehydration on the protonic, or ionic, motion throughout the hydration
layers surrounding soluble proteins. The amount of water around each 
protein is crucial for free protonic motion around the molecule. By
dehydrating these water layers, a level is being reached when protonic
motion becomes awkward and stops, and so does also, simultaneously, 
the enzymatic activity
of the protein molecule. On re-hydrating the molecule, protonic
motion becomes possible again and this, in turn, causes also the onset
of enzymatic activity of the protein molecule. This experiment shows 
that free ionic motion around soluble protein molecules is crucial
for their enzymatic activity. 
\vspace{0.5cm}

\noindent {\bf III Physical Aspects and Biological Significance} \\

The main objective of the present article is to draw several physical
conclusions from the phenomenon of chiral interaction 
in macro-chiral devices
which are quite different from the
regular rules that exist in classical physics. The source of these 
differences arises from the presence of chirality as a major physical
object instead of the Newtonian mass point that plays a basic role in
classical mechanics and is also of ideal spherical symmetry, that is, 
completely achiral. From these conclusions analogies
can be drawn for the function of molecular chiral systems which may
well be of considerable significance in molecular biology. \\

The first conclusion concerns the symmetry operation of time-reversibility
that exists in many examples of classical physics. In the case of chiral
devices time-reversibility does not exist. The windmill, for example,
rotates about its axis in a given direction due to its chiral design.
Upon reversing time, the rotational velocity changes its direction,
so that it rotates in the opposite direction. This cannot happen
mechanically, since there is no mechanism in a windmill that can 
rotate it backward. What is rotating in 
an opposite direction is the {\bf mirror-image} 
of the given windmill but not the given windmill itself. 
The meaning of this mode of symmetry operation is that a windmill
is time-irreversible, but it obeys space-time invariance.
Let us now express the space-time inversion by $P$ and $T$,
respectively: \ then a windmill does obey the PT-invariance transformation.
The same is true for all the examples given here of macro-chiral devices. 
The same is also true for the protein
molecule example. This rule of PT-invariance (or CPT invariance) is 
recognized in physics due to the presence of a spin in quantum mechanics,
but is absent in classical physics because the concept
of structural chirality in physics it is largely ignored. 
This concept appears much more in 
chemistry due to the presence of many chiral molecules in organic 
compounds, but chirality is mostly regarded and treated in chemistry
in terms of shape, rather
than in its physical properties and contents. For this reason the concept 
of CI has so far
been largely overlooked in researches concerning chirality. \\

These space-time symmetry operations for chiral devices contain also a 
certain aspect of practicality.  This is in contrast to their 
presence in the
domain of elementary particles in physics. From this view point any  
time-reversible process is almost completely useless from any aspect of 
practicality. For instance, any machine operation that produces 
a certain function or object,
or any information transfer process are completely time-irreversible. 
These
include also biomolecular functions such as enzymatic activity and other
processes which are totally time-irreversible. For such reasons 
of practicality
the function of chiral devices or molecules is of special significance
in comparison to the time reversible phenomenon that appears in many 
physical operations that involve the presence of the Newtonian mass
point. \\

The next consideration involves the mode of selection where only 
{\bf one} direction of rotation is excited by CI, whereas the opposite
direction remains largely inactive. Judging it from a thermodynamical
aspect, what is happening here is that only one half of the energy 
that can be activated by the device is excited by CI 
whereas the other half 
remains inactive. On a molecular level this means that only
one half of the energy states of the system are populated by 
CI become active, whereas
the other half remains empty. In other words the system does not readily
reach thermal equilibrium. This conclusion is of very substantial and
significant meaning for living systems because reaching
thermal equilibriuim means death. \\

Another way to look at this effect is from the view-point of ergodicity.
This concept was introduced by Boltzmann about a century ago and it regards
the mode of approaching thermal equilibrium of a single particle. 
This is done in a 
process of time average instead of an ensemble statistical average. In 
view of this, the average velocity of such a particle in any given 
direction approaches zero as a function of time. This is not the case
if, for instance, the average angular velocity of a windmill is regarded
as a function of time. This is, actually, true for any effect of CI when
averaged as a function of time. The selection of one direction of 
motion out of two possible ones, which is typical of CI, makes its mode
of motion to become a non-ergodic entity, which again causes it to 
avoid thermal equilibrium. This property of CI on a microscopic level
is, apparently, one of the most crucial advantages that chirality, or 
CI, does contribute to molecular biology. It does postpone thermal 
equilibrium, or death, for a considerable length of time, so that 
the biological 
function of these molecules can go on and not be affected by 
approaching thermal equilibrium. \\

In this context it is interesting to mention also Schr\"{o}dinger 
who became
interested in the phenomenon of life and wrote a book ``What is Life?''
in 1944$^{14}$. His main conclusion in this book was: ``It feeds on 
negative entropy'', and this is exactly what CI is performing in its
mode of selecting only one direction of motion. It is thus reducing the 
entropy of the system. \\

In relation to the phenomenon of non-ergodicity it is also important
to mention its relevance to the process of evolution, which is 
crucial in biology.
It is reasonable to deduce that systems that reach readily thermal 
equilibrium never undergo the process of evolution and remain 
basically unchanged forever. Non-ergodic molecular systems have a 
better chance to undergo evolutionary changes. 
\vspace{0.5cm}

\noindent {\bf IV \  Discussion and Conclusions} \\

Another aspect of CI regards the nature of this effect, as well as the
amount of energy that is involved in such a process. In discussing this 
case it is not relevant to consider macro-chiral devices and our main
concern is CI of biomolecular systems. Unfortunately, our
knowledge, at present, of this effect is very limited and this is mainly 
because of 
the small amount of energy involved in this effect, being, in fact, 
subthermal in size$^{5-6,10}$, which is quite difficult to observe 
experimentally. This may evoke criticism as to its possible significance.
Such a criticism is rather common among scientists who tend to attribute
significance to energy according to its size. What may be much more 
significant than the amount of energy involved in a process, is its
quality, or degree of sophistication. This is particularly so in complex
systems such as certain biomolecules, proteins for example. The feature
of time-irreversibility of CI does contribute a degree of sophistication.
In addition to this there exist quite a few examples of highly sophisticated
modes of energy which require rather minute quantities of energy. 
For instance,
an information transfer process requires a high degree of sophistication in 
wave modulation and its size of energy is relatively small. In comparison,
boiling a kettle of water requires much more energy, but 
what is its degree of sophistication? Another example is the small 
amount of energy required to switch on and off a much larger source
of energy. This example can be regarded as a mode of 
control mechanism energy 
which may also be the significance of CI in biology. 
Another, rather cruel example,
concerns the magnitude of energy change that occurs over a short time
interval during which a creature ceases to live. The change in energy
is quite small but its significance is impressive. In these examples
and many others, the amount of energy involved in their performance
is of little interest, but their main effect is in their degree of 
sophistication. \\

It is too early now to attempt to specify any definite mode of 
sophisticated performance of CI on a biomolecular level. Such effects
have to be studied further in order to become better
understood. The experiment of Careri et al.$^{13}$ provides for a 
supporting evidence for the significance of CI in the enzymatic 
activity of proteins. It is quite reasonable to assume that in 
biology, or in any living substance, the phenomenon of existence of 
such modes of highly sophisticated and
low energy signals may have an important function 
in its life process. \\

In conclusion, let us mention again the significance and importance 
of the phenomenon
of chirality in biology, in particular the features of chiral 
interaction (CI) that differ largely from those of classical
physics that do not contain chiral structure in their interactions.
These include the PT-invariance of chiral interaction, which causes 
it to be time-irreversible. The selectivity nature of CI
by preferring one mode of motion out of two possible ones, enables CI
to become non-ergodic, which is a crucial element in life processes
and biological evolution. 
\pagebreak

\begin{center}
{\bf References}
\end{center}

\begin{enumerate}
\item F. Arago, ``Memoires de la Classe des Sciences Math. et Phys. de
l'Institut Imperial de France'', Part 1, p. 93 (1811).
\item J.B. Biot, ``Memoires de la Classe des Sciences Math. et Phys. de
l'Institut Imperial de France, Part 1, 1 (1812).
\item L. Pasteur, Ann. Chim. {\bf 24}, 457 (1848).
\item W.T. Kelvin, ``Baltimore Lectures'', C.J. Clay \& Sons, London
(1904).
\item G. Gilat, Chem. Phys. Lett. {\bf 121}, 9 (1985).
\item G. Gilat, Mol. Eng. {\bf 1}, 161 (1991).
\item G. Gilat, ``The Concept of Structural Chirality'', in ``Concepts
in Chemistry'', Ed. D.H. Rouvray (Research Studies Press and Wiley \&
Sons, London, New York, 1996) p. 325.
\item G. Gilat, J. Phys. A.{\bf 22}, p. L545 (1989) ibid Found. Phys. Lett.
{\bf 3}, 189 (1990).
\item G. Gilat and L.S. Schulman, Chem. Phys. Lett. {\bf 121} 13 (1985).
\item G. Gilat, Chem. Phys. Lett. {\bf 125}, 129 (1986).
\item G. Gilat, to appear in the Proceedings of a Conference on
``Biological Homochirality'', Serramazzoni, Italy, September 1998.
\item H. Tschersche in ``Biophysics'', Eds. W. Hoppe, W. Lohmann, 
H. Markl \& H. Ziegler (Springer Verlag, Berlin 1983) p. 37.
\item G. Careri, A. Giasanti and J.A. Rupley, Phys. Rev. {\bf A37},
2763 (1988).
\item E. Schr\"{o}dinger, ``What is Life'', Cambridge University 
Press, Cambridge, 1944.
\end{enumerate}
\vspace{0.5cm}


\begin{figure}
\centerline{\epsfig{file=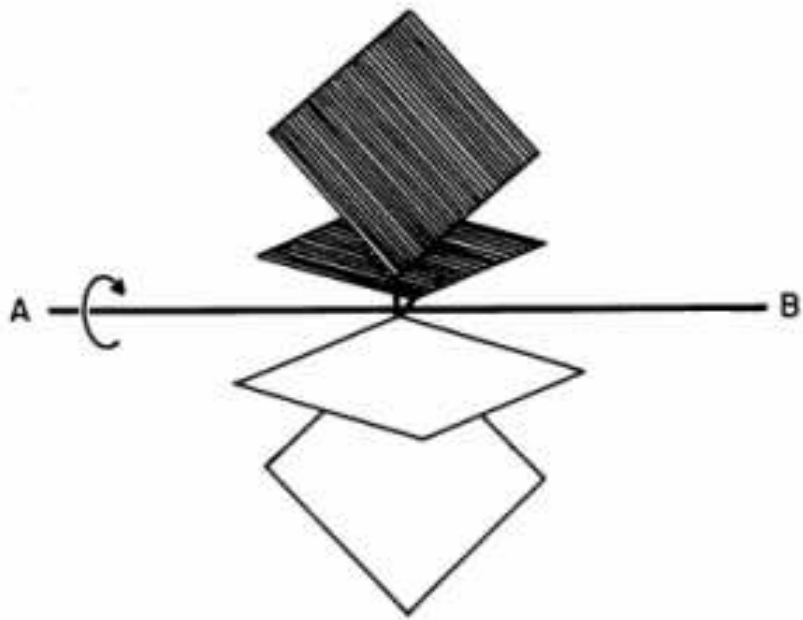, width=\columnwidth}
}

\caption{ A variant of Crookes' radiometer is an example
of chiral interaction (CI). The asymmetry in the optical 
absorption coefficient
between the black and the silver blades generates a temperature difference
between them when light is shining at the device. This expands the air 
close to the black blade which, in turn, pushes it around the axis AB in
the preferred direction towards the black vane. This is an 
example of physical
rather than of a geometric chirality}
\label{fig}
\end{figure}
\end{document}